\documentclass[fleqn,twoside]{article}

\usepackage{espcrc2}                                 % Elsevier style
\usepackage{graphicx}
\usepackage{amsmath,amsfonts}
\usepackage[numbers,sort&compress]{natbib}           % for [1,2,3] --> [1-3] 
\graphicspath{{./Figures/}}                         % location for color fig.
\newcommand{\op}{\langle\bar{\psi}i\gamma_{5}\tau^{3}\psi\rangle}  % order para.
\newcommand{\bkplane}{\mbox{$(\beta, \kappa)$-plane }}             % b-k-plane
\newcommand{\etal}{\textit{\mbox{et al.\ }}}                       % et. al.
\title{%
\thispagestyle{empty} \vspace{-25mm} 
\begin{flushright}
     \small HU--EP--03/54\\
     \small September 2003
\end{flushright}
\vspace{8mm}
The Aoki phase for $N_f=2$ Wilson fermions revisited
\thanks{Talk presented by A.~Sternbeck
at the International Symposium Lattice 2003, Tsukuba, Japan%
.}}

\author{A.~Sternbeck\address[HU]{Institut f\"ur Physik, 
        Humboldt-Universit\"at zu Berlin, D-12489 Berlin, Germany},
        E.-M.~Ilgenfritz\addressmark[HU],
        W.~Kerler\addressmark[HU], M.~M\"uller-Preussker\addressmark[HU] 
        and
        H.~St\"uben\address{Konrad-Zuse-Zentrum 
        f\"ur Informationstechnik Berlin, D-14195 Berlin, Germany}}

\begin{document}

\begin{abstract}
  We report on a numerical reinvestigation of the Aoki phase in full lattice 
  QCD with two flavors of unimproved Wilson fermions. For zero temperature
  the Aoki phase can be confirmed at inverse coupling $\beta=4.0$ and 
  $\beta=4.3$, but not at $\beta=4.6$ and $\beta=5.0$. At non-zero temperature
   the Aoki phase was found to exist also at $\beta=4.6$.
\end{abstract}

% typeset front matter (including abstract) and HU-EP Number
\maketitle

%-------------------------------------------------------------------------------
\section{The Aoki phase}

In studies of chiral symmetry breaking it is desirable to start
from an (almost) chirally invariant formulation of lattice fermions. 
At present this is best realized by using fermions satisfying the 
Ginsparg-Wilson relation, e.g.~using the overlap operator, or by working
with the domain wall framework \cite{GWferm}. In such approaches Wilson 
fermions are still important as an input. 

\begin{figure}[htb]
  \centering
  \includegraphics[height=4.6cm]{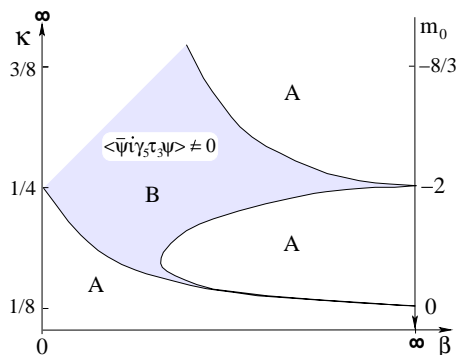}
  \vspace{-0.7cm}
  \caption{Phase structure proposed by Aoki \etal\!\!. 
    The shaded region $B$ is the phase
    where flavor and parity are spontaneously broken. 
    Both symmetries are conserved in regions $A$.}
  \label{fig:proposed_phase_diagram}
\end{figure}

For the Wilson-Dirac operator (which breaks chiral invariance explicitly) in
$N_f=2$ lattice QCD Aoki \cite{Aoki3} has predicted that in a certain 
parameter range there is a phase in which parity and flavor symmetry are 
both spontaneously broken. It is separated from an unbroken phase by a 
line of second order phase transition on which the pion states are 
expected to become massless \cite{Aoki4}. 
In Fig.~\ref{fig:proposed_phase_diagram} the proposed phase 
diagram is sketched in the \bkplane\!.

In a recent paper \cite{GolSham} it has been pointed out that the features of 
the Aoki phase are of relevance for locality and for restoration of chiral
invariance in quenched and full QCD with Ginsparg-Wilson and domain wall
fermions. Accordingly, the region of the Aoki phase has to be avoided in
such investigations in order not to spoil physical reliability.

In the light of this it is important to have precise information about the
region where the Aoki phase really occurs. In the past it has been questioned 
\cite{Sharpe1} whether the Aoki phase exists also at larger $\beta$-values,
i.e.~towards the continuum. However, previous investigations of this issue
did not lead to a unique answer. Therefore we have addressed this 
problem again \cite{BQCD}. 

Here we report on results showing that most likely the Aoki phase does not 
exist for \mbox{$\beta\ge4.6$} at zero temperature. Furthermore, we 
present data for finite temperature indicating such broken phase 
still at $\beta=4.6$ and connecting it to the finite temperature 
phase transition.
\vspace*{-4mm}

%------------------------------------------------------------------------------
\vspace*{-8mm}
\section{Simulation details}

\begin{figure}[htb]
  \hspace{-0.4cm}
  \mbox{\includegraphics[height=6.5cm]{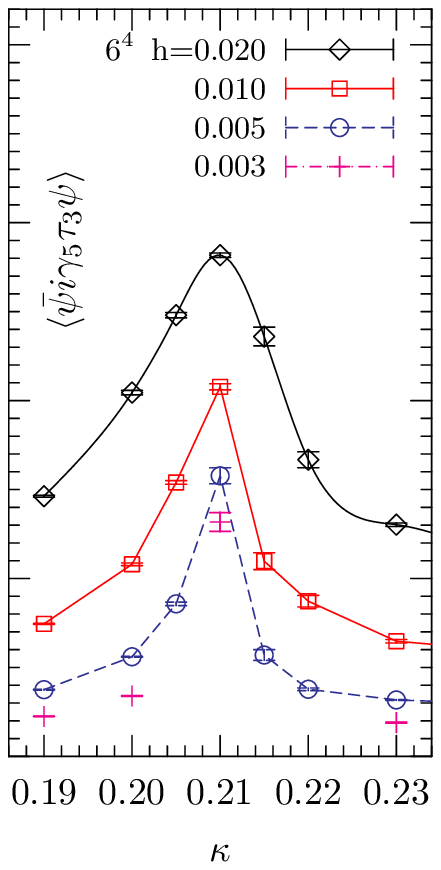}\hspace{-0.8cm}%
    \includegraphics[height=6.5cm]{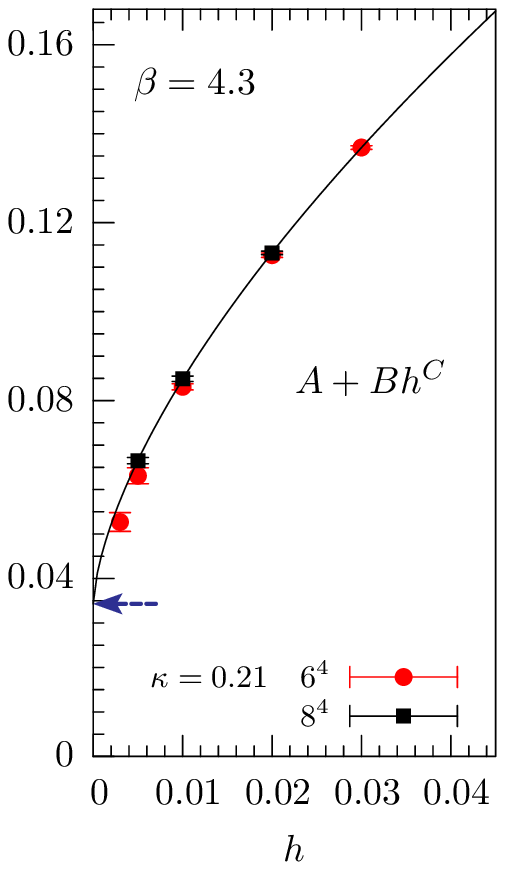}}
  \vspace{-10mm}
  \caption{$\op$ for $\beta=4.3$ as a function of $\kappa$ 
           at several values of $h$ (left, the lines are spline 
           interpolations to guide the eye) and as a function of $h$ 
           at $\kappa=0.21$ (right, the line represents a fit described 
           in the text). The arrow marks the extrapolated value of 
           the order parameter.}
  \vspace{-1mm}
  \label{fig:extrapolation43}
\end{figure}

To investigate the Aoki phase we have simulated lattice QCD with two flavors 
of unimproved Wilson fermions using the Hybrid Monte Carlo algorithm.  An 
explicitly symmetry-breaking source term was added to the fermion matrix 
$M_W$, 
\begin{equation}
  \label{eq:fermionmatrix}
  M(h)=M_W+ h i\gamma_5\tau^3,
\end{equation}
since without the added term the order parameter $\op$ would always be zero 
on a finite lattice. Then $\op$ was measured varying the lattice size 
$V$ and the (non-zero) $h$-values. The order parameter $\op_{h=0}$ 
monitoring spontaneous symmetry breaking is obtained 
by taking the double limit in the following order
\begin{equation}
\label{eq:lim_h_V}
  \op_{h=0}=\lim_{h\rightarrow 0}\lim_{V\rightarrow \infty}
  \langle\bar{\psi} i \gamma_{5} \tau^{3}\psi\rangle \,.
\end{equation}

For zero temperature the simulations were performed on lattices 
ranging from $4^4$ to $12^4$ at $\beta$-values 4.0, 4.3, 4.6, and 5.0, 
with $\kappa$ and $h$ in the intervals $0.15\le\kappa\le0.28$ 
and $0.003\le h\le0.04$, respectively. For each $\beta$ and each lattice
size (starting on $6^4$) the physically relevant values of 
$\kappa$ were determined first, namely the region where $\op$ has 
a peak at finite $h$. 

\begin{figure}[htb]
  \hspace{-0.7cm}
  \mbox{\includegraphics[height=6.5cm]{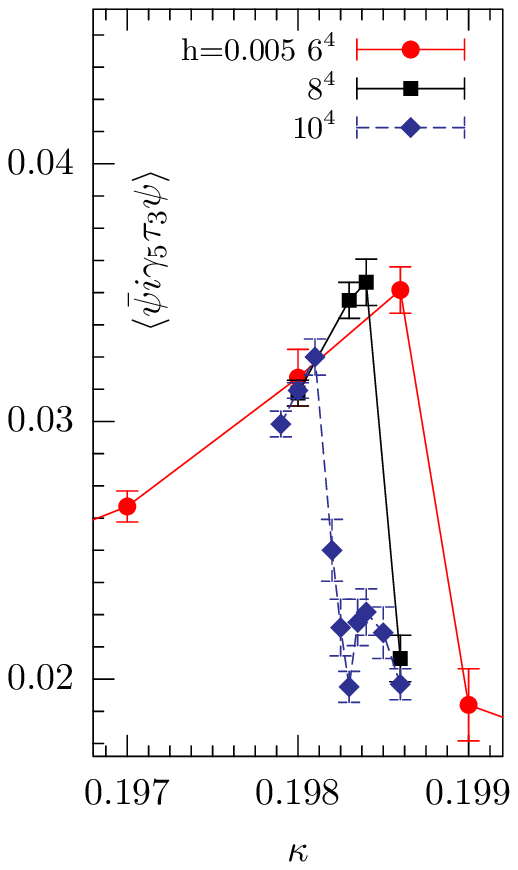}\hspace{-0.2cm}%
        \includegraphics[height=6.5cm]{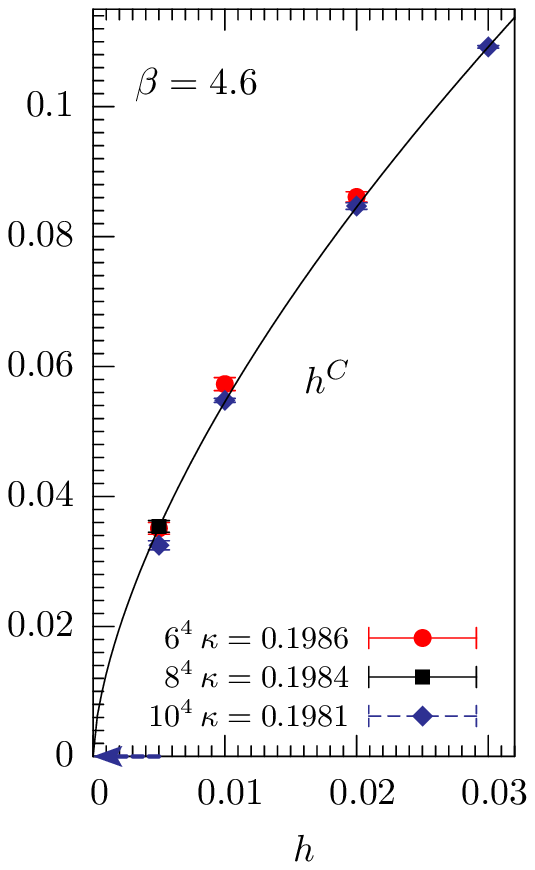}}
  \vspace{-14mm}
  \caption{$\op$ for $\beta=4.6$ at $h=0.005$ as a function of $\kappa$, 
           for different lattice sizes (left, the lines are drawn to guide 
           the eye), and as a function of $h$ for $\kappa$ changing 
           from 0.1986 to 0.1981 (right, the line represents a fit 
           described in the text). The arrow marks the extrapolated value of
           the order parameter.}
  \vspace*{-5mm}
\label{fig:extrapolation46}
\end{figure}

This is shown on the left hand sides of Figs.~\ref{fig:extrapolation43} and 
\ref{fig:extrapolation46}. One sees that for $\beta=4.3$ and $\beta=4.6$ the
crucial values are near $\kappa=0.21$ and $\kappa=0.1981$, respectively. 
At these values we have increased the lattice sizes until the measurements 
for different sizes agreed within errors. In this way we achieved that 
we can treat our largest lattices as infinitely large.

%----------------------------------------------------------------------------
\section{The extrapolation {\boldmath $h \rightarrow 0$} }

As described in Ref.~\cite{BQCD} a reasonable ansatz to extrapolate the 
data to $h=0$ is 
\begin{equation}
  f(h) = A + Bh^{C} + \ldots  \; .
  \label{eq:ansatz}
\end{equation}
With $C$ as a free parameter it describes the data well. Indeed, the 
parameter of interest, $A$, is then robust against the introduction of 
correction terms linear and quadratic in $h$.

\begin{figure}[t]
  \hspace{-0.6cm}
  \includegraphics[width=0.5\textwidth]{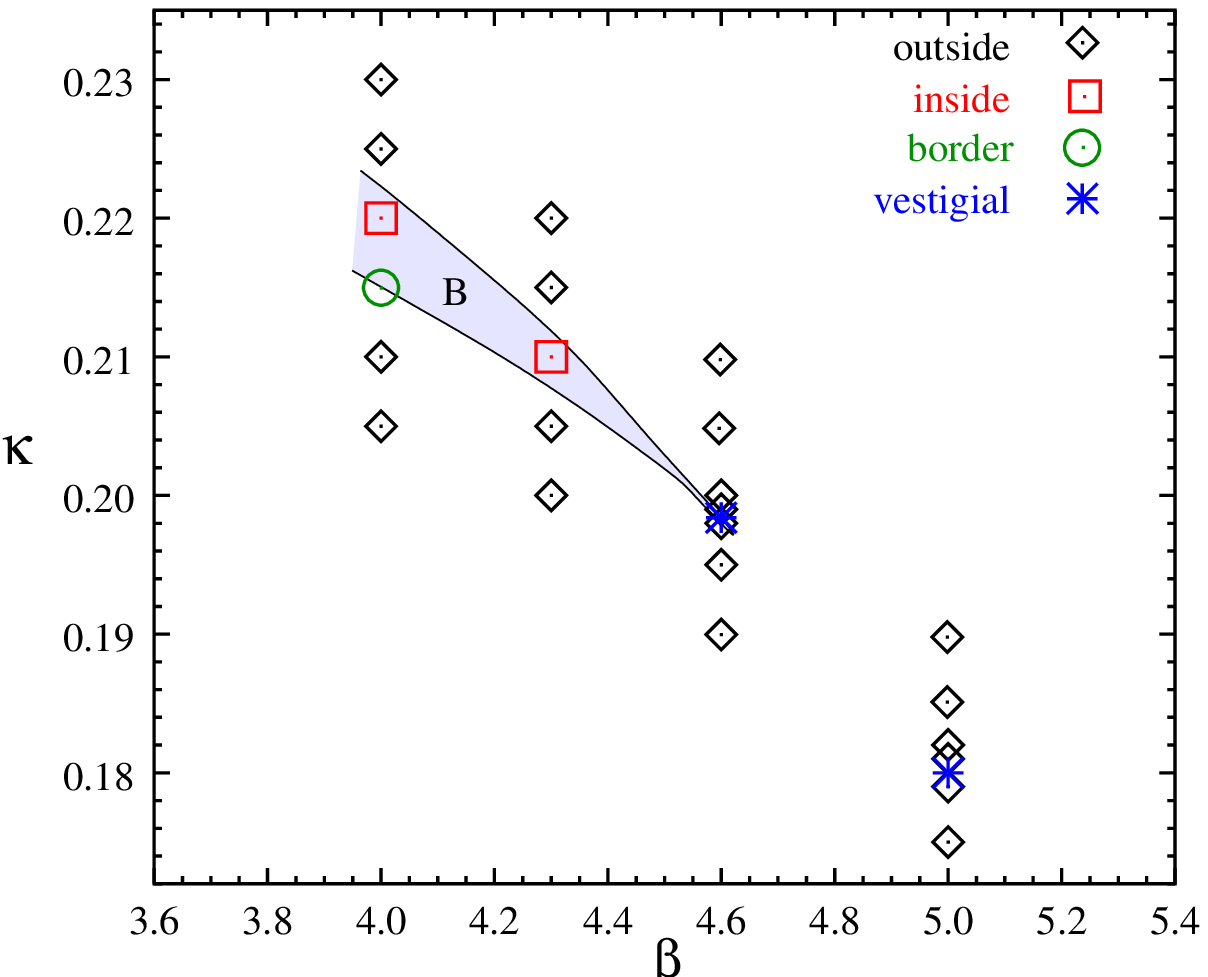}
  \vspace*{-12mm}
  \caption{The area of the phase diagram studied in Ref.~\cite{BQCD}.  
           Squares indicate that $\op_{h=0}$ is finite, diamonds indicate 
           that $\op_{h=0}$ vanisches, and stars denote points where 
           $\op_{h=0} \ne 0$ is uncertain.  The lines sketch the limits 
           of the Aoki phase.  The data indicate that the point 
           ($\beta,\kappa$)=($4.0,0.215$) (circle) is very close to the 
           lower critical line \cite{BQCD}.}
 \label{fig:phasediagram_results}
\end{figure}

At $\beta=4.3$ the parameter $\op$ has a non-vanishing extrapolation, 
while at $\beta=4.6$ the best fit gives $A=0$. Combined with similar 
simulation results at $\beta=4.0$ and $\beta=5.0$ (see \cite{BQCD}) we 
confirm the Aoki phase at $(\beta,\kappa)=(4.0,0.22)$ and $(4.3,0.21)$. 
The fit parameters $B$ and $C$ turn out to agree for different $\beta$
within errors (the values are $B\approx1$ and $C\approx0.65$). 
At $\beta=4.6$ the 
ansatz also works, but results in a vanishing order parameter.  Also at 
$\beta=5.0$ no finite order parameter was found \cite{BQCD}. This suggest 
that the Aoki phase ends somewhere near $(\beta,\kappa)=(4.6,0.1981)$ as 
shown in Fig.~\ref{fig:phasediagram_results}. The shift in $\kappa$ with 
increasing lattice size (see Fig.~\ref{fig:extrapolation46}) appears to 
be related to the end of the symmetry-breaking phase.

%-------------------------------------------------------------------------------
\section{Outlook for finite temperature}

\begin{figure}[t]
  \vspace{-3mm}\hspace{-1cm}
  \mbox{\includegraphics[height=6.5cm]{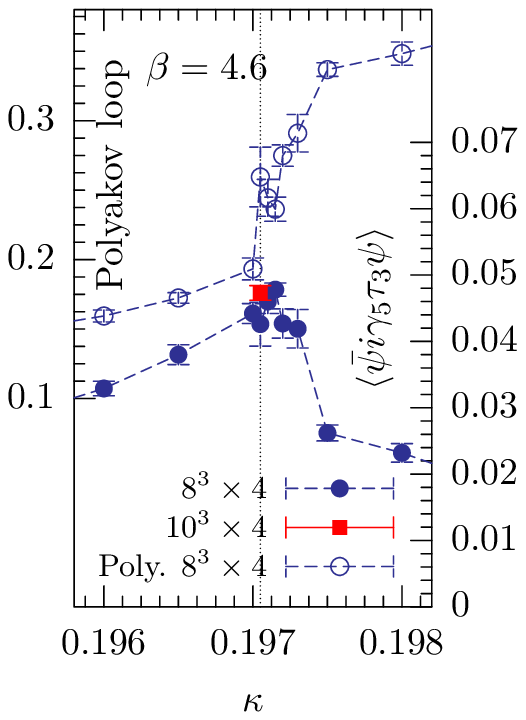}\hspace{-0.75cm}%
    \includegraphics[height=6.5cm]{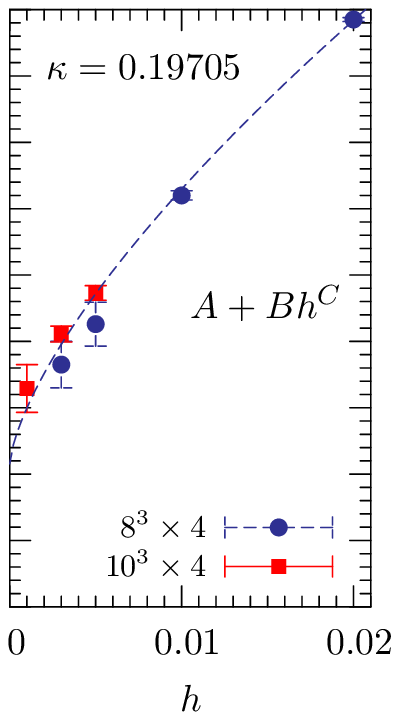}}
  \vspace{-14mm}
  \caption{The parameter $\op$ and the Polyakov loop as function of 
           $\kappa$ on a $8^3\times4$ and a $10^3\times4$ lattice at 
           $h=0.005$ and $\beta=4.6$ (left, the lines are drawn to guide 
           the eye). At $\kappa=0.19705$ we obtained $\op\ne0$ from an 
           extrapolation explained in the text (right).} 
\label{fig:extrapolation46finT}
\end{figure}

We have also performed simulations for the finite-temperature case.
In contrast to zero temperature we find the Aoki phase at $\beta=4.6$ 
as shown in Fig.~\ref{fig:extrapolation46finT}.  One can see that the 
Polyakov loop (open symbols) steeply rises near $\kappa=0.19705$ where 
$\op$ (filled symbols) extrapolates to a finite value. This means that 
the finite-temperature phase transition occurs close to the Aoki phase.  
We were not able to detect a separation of this transition from the 
Aoki phase. This is in contrast to Ref.~\cite{Aoki14} where a 
scenario with separated transition lines is proposed. 

%---------------------------------------------------------------------------
\section*{Acknowledgements}

All simulations were done on the Cray T3E at Konrad-Zuse-Zentrum f\"ur 
Infor\-mations\-technik Berlin. A.~S.~would like to thank the DFG-funded 
graduate school GK~271 for support.

%--------------------------------------------------------------------------
% Bibliography:

%=============================================================================
\end{document}